\documentclass[10pt,british,english,aps,superscriptaddress,preprintnumbers,amsmath,nofootinbib,amssymb,altaffilletter]{revtex4}
\usepackage[T1]{fontenc}
\usepackage{xcolor}
\usepackage{babel}
\usepackage{float}
\usepackage{amsmath}
\usepackage{amssymb}
\usepackage{graphicx}
\usepackage{esint}
\usepackage{braket}

\usepackage[unicode=true,pdfusetitle,
 bookmarks=true,bookmarksnumbered=false,bookmarksopen=false,
 breaklinks=false,pdfborder={0 0 1},backref=false,colorlinks=false]
 {hyperref}

\makeatletter
\@ifundefined{textcolor}{}
{%
 \definecolor{BLACK}{gray}{0}
 \definecolor{WHITE}{gray}{1}
 \definecolor{RED}{rgb}{1,0,0}
 \definecolor{GREEN}{rgb}{0,1,0}
 \definecolor{BLUE}{rgb}{0,0,1}
 \definecolor{CYAN}{cmyk}{1,0,0,0}
 \definecolor{MAGENTA}{cmyk}{0,1,0,0}
 \definecolor{YELLOW}{cmyk}{0,0,1,0}
}

\usepackage{babel}
\usepackage{babel}
\usepackage{babel}
\usepackage{babel}

\usepackage{babel}
\usepackage{breakurl}

\@ifundefined{textcolor}{}{%
 \definecolor{BLACK}{gray}{0}
 \definecolor{WHITE}{gray}{1}
 \definecolor{RED}{rgb}{1,0,0}
 \definecolor{GREEN}{rgb}{0,1,0}
 \definecolor{BLUE}{rgb}{0,0,1}
 \definecolor{CYAN}{cmyk}{1,0,0,0}
 \definecolor{MAGENTA}{cmyk}{0,1,0,0}
 \definecolor{YELLOW}{cmyk}{0,0,1,0}
}

\usepackage{babel}
\usepackage{epsfig}\usepackage{mathrsfs}\usepackage{euscript}

\renewcommand{\[}{\begin{equation}}
\renewcommand{\]}{\end{equation}}
\def\beq{\begin{equation}}
\def\eeq{\end{equation}}
\newcommand{\be}{\begin{eqnarray}}
\newcommand{\ee}{\end{eqnarray}}

\renewcommand{\texttt}{{}}

\def\bs{\begin{subequations}}
\def\es{\end{subequations}}

\def\Cc{\mathcal{C}}

\def\Fc{\mathcal{F}}

\def\Kc{\mathcal{K}}

\def\Mc{\mathcal{M}}

\def\Oc{\mathcal{O}}
\def\Pc{\mathcal{P}}

\def\Rc{\mathcal{R}}

\def\Wc{\mathcal{W}}
\def\Zc{\mathcal{Z}}

\newcommand{\tia}[1]{}

\newcommand{\bea}{\begin{eqnarray}}
\newcommand{\eea}{\end{eqnarray}}
\newcommand{\beas}{\begin{eqnarray*}}
\newcommand{\eeas}{\end{eqnarray*}}
\newcommand{\bal}{\begin{aligned}}
\newcommand{\eal}{\end{aligned}}

\def\({\left(}
\def\){\right)}

\newcommand{\LF}{\left(}
\newcommand{\RF}{\right)}
\newcommand{\LT}{\left[}
\newcommand{\RT}{\right]}

\newcommand{\pd}{\partial}

\newcommand{\const}{\mathrm{const}}

\makeatother

\begin{document}

\title{Non-local Starobinsky inflation in the light of future CMB}

\author{K. Sravan Kumar}
\email{sravan@ubi.pt,korumilli@sustc.edu.cn}

\selectlanguage{british}%

\address{Department of Physics, Southern University of Science and Technology (SUSTech), Shenzhen 518055, P.R. China}
\selectlanguage{english}%

\author{Leonardo Modesto}
\email{lmodesto@sustc.edu.cn}

\selectlanguage{british}%

\address{Department of Physics, Southern University of Science and Technology (SUSTech), Shenzhen 518055, P.R. China}
\selectlanguage{english}%

\begin{abstract}
Analytic infinite derivative (AID) non-local quadratic curvature gravity in Weyl basis is known to be ghost free, superrenormalizable or finite and perturbatively Unitary and as such it is Ultra-Violet (UV) complete.  Recently $R+R^2$ ("Starobinsky") inflation was successfully embedded in AID non-local gravity and the corresponding observables were computed. Here in this paper, we derive the  form factors compatible within near de Sitter aproaximation and prove that the theory must contain a scalaron that drives inflationary expansion. Further more we consider the form factors (AID non-local operators) proposed by Tomboulis in hep-th/9702146 and compute the corresponding predictions of tensor to scalar ratio and tensor tilt $\LF n_t,\,r \RF$ where the scalar tilt remains the same as the local Starobinsky model. Anticipating future CMB probes such will be able to test non-local Starobinsky inflation we constrain the scale of non-locality to be $10^{14} GeV\lesssim\Mc\lesssim  5\times 10^{14} GeV$ and $10^{-7}\lesssim r \lesssim 0.07$ for different form factors. We found that it possible to have a blue or red tensor tilt $\LF n_t\gtrless 0 \RF$ depending on the scale of non-locality and the form factor. We also comment on Higgs inflation in non-local context. 
\end{abstract}
\maketitle

\section{Introduction}

\label{sec.intro}

Cosmic inflation is the most successful theory of early Universe \cite{Starobinsky:1980te,Guth:1980zm,Linde:1981mu}.
The first model of inflation is based on simple one parameter extension
of General relativity (GR) with an additional $R^{2}$ term (where
$R$ is the Ricci scalar) in the action and it is known as ``Starobinsky
`` inflation \cite{Starobinsky:1980te}. With the latest Cosmic microwave background measurements by Planck
satellite $R^{2}$ inflation stands out to be in the central spot
in the plane of key inflationary predictions \footnote{which emerge from the theory of quantum fluctuations
	 imprinted as the temperature fluctuations $\frac{\Delta T}{T}\sim10^{-5}$
	in the CMB which are the seeds to the large scale structure formation
	\cite{Mukhanov:1981xt,Mukhanov:1990me,Mukhanov:2013tua}.} such as scalar tilt $n_{s}$ and the ratio of tensor to scalar power
spectrum $r$ which reads as 

\begin{equation}
n_{s}=1-\frac{2}{N}\,,\quad r=\frac{12}{N^{2}}\label{nsrR2local}
\end{equation}
where '$N\sim50-60$' is the required number of $e-$foldings from
the end to the beginning of inflation consistent to resolve horizon
and flatness problems. The current constraints imply $n_{s}=0.968\pm0.006$
and $r<0.09$ \cite{Ade:2015lrj,Array:2015xqh} where no conclusive
evidence on power spectrum of primordial gravitational waves has been
found yet. However, $R^{2}$ inflation is so far well consistent with
$n_{S}\sim0.962$ and $r\sim0.004$ for $N\sim55$. Alternatively,
inflationary scenarios based on scalar field dynamics were proposed
which are commonly known as ' chaotic inflation ' \cite{Linde:1982uu,Linde:1983gd}.
Scalar field models heavily rely on the shape of scalar potential
during inflation which is essentially needs to be flat to match with
current observations. Among these, several models were heavily constrained
or ruled out by 2015 Planck data \cite{Ade:2015lrj,Linde:2014nna,Martin:2015dha}.
The first main reason for greater interest of studying several inflationary
scenarios is that it gives the best opportunity to test theories of
gravity promising UV completion. This propelled physicists to construct
models within string theory/supergravity
(SUGRA). There has been plethora of inflationary models exist in the
literature based on several modifications of matter or gravity sector
inspired from settings of string theory \cite{Martin:2013tda,SravanKumar:2018reb}. The
second interesting reason inflation is so convincingly abridge the
cosmology with present day known particle physics through the process
of reheating where the scalar mode\footnote{which is usually named as inflaton in scalar field models or scalaron
	in modified gravity models.} that drives expansion settles to a (true) vacuum leading particle
production \cite{Kofman:1994rk,Kofman:1997yn,Linde:2005ht,SravanKumar:2018reb}. Therefore, several particle physics models were proposed within
grand unified theories (GUT)\cite{Shafi:1983bd,Linde:2005ht}. A more
interesting development happened when the standard model Higgs\footnote{which was found in 2012 LHC run \cite{Aad:2012tfa}.}
(with non-minimal coupling to $R$) was proposed to be candidate for
inflaton and it was shown to give exactly same predictions of $R^{2}$inflation\footnote{However, there were some theoretical arguments that favors $R^{2}$
	inflation over inflation \cite{Salvio:2015kka}.} \cite{Bezrukov:2007ep,Bezrukov:2011gp}. The proposals of observationally
degenerate models of inflation emphasized the vital importance to
realize inflation within an attractive scopes of theories which are
likely to be UV complete. For example, there were attempts to realize $R^{2}$
and Higgs inflation in string theory/SUGRA\footnote{Also several more inflationary constructions in string theory/SUGRA that are consistent with recent data
	be found in \cite{Burgess:2013sla}.}  \cite{Ketov:2010qz,Ferrara:2010yw}.

The original motivation for the Starobinsky inflation is from the early attempts of quantum gravity by Stelle and the well known issue of conformal anomaly  \cite{Stelle:1976gc,Stelle:1977ry,Duff:1993wm}. Stelle has shown that $R^2$ modification of gravity including a quadratic in Weyl square term is renormalizable with a price of a massive tensor ghost that spoils the Unitarity. However, it is probably the right motivation from quantum gravity would be the reason for Starobinsky model being stood as the best fit with respect to the current observations \cite{Ade:2015lrj}. The long shot of having a UV complete theory of quantum gravity is well fulfilled by the AID non-local theories. This is the most significant step from Stelle's 4th order theory of gravity which was renormalizable and non-unitary. Non-local gravity theories were initiated in \cite{Kuzmin:1989sp,Krasnikov:1987yj,Tomboulis:1997gg} 
and extensively progressed in recent years showing these are unitary (ghost-free) and  super-renormalizable or finite in the framework of QFT  around maximally symmetric spacetimes
\cite{Biswas:2005qr,Biswas:2011ar,Modesto:2011kw, Briscese:2013lna,Biswas:2013cha,Calcagni:2014vxa,Alexander:2012aw,Modesto:2014lga,Modesto:2015lna,Tomboulis:2015esa,Tomboulis:2015gfa,Koshelev:2017ebj}. The issue of causality, initial conditions and non-perturbative degrees of freedom in these gravitational theories is further investigated in recent works \cite{Calcagni:2018lyd,Calcagni:2018gke,Briscese:2018oyx,Buoninfante:2018mre}.

The UV completion of $R^{2}$ was shown to occur in a  non-local theory of gravity containing infinite higher derivative terms like $R\mathcal{F}_{R}\left(\Box\right)R$
and $W_{\mu\nu\rho\sigma}\mathcal{F}_{W}\left(\Box\right)W^{\mu\nu\rho\sigma}$
(where $W$ is the Weyl tensor) \cite{Koshelev:2016xqb}.
The new feature of this model is
that it modifies the tensor power spectrum leading to arbitrary prediction
of $r<0.07$. In recent study \cite{Koshelev:2017tvv}, Starobinsky inflation was shown to be an attractor in this non-local gravity theory and also a careful study of perturbations revealed that the scalar power spectrum and its tilt remain the same as local $R^2$ inflation where as the tensor power spectrum and the tilt gets modifications as such we do modify the single field inflationary consistency relation \cite{Mukhanov:1990me,DeFelice:2010aj}. 

The paper is organized as follows. In Sec.~\ref{gen-sec} we review the Starobinsky's local $R^2$ theory of gravity and quasi-de Sitter (dS) solution. Then we present the non-local extension Starobinsky theory where we introduce AID operator for quadratic Ricci scalar term and also add Weyl square term with an AID operator inbetween. We describe the properties of the theory such as super renormalizability and perturbative Unitary (ghost free). We discuss the most general solution of the theory which was found for conformally flat backgrounds \cite{Koshelev:2017tvv}. In Sec.~\ref{sec.starouv} we compute the second order action of the theory for spin-0 and spin-2 parts in near dS approximation. We consistently find the general structure of AID operators of Ricci suqare and Weyl square terms. Later we show that the general solution of the theory around conformally flat metrics essentially require a propagating scalar degree of freedom (scalaron). In Sec.~\ref{sec.staroin} we review the inflationary perturbations of non-local $R^2$ model and update the predictions with the form factors found in Sec.~\ref{sec.starouv}. Considering the entire functions that define the UV properties of the theory proposed in \cite{Tomboulis:1997gg} we compute the inflationary observables of non-local $R^2$ model and discuss against the future forecasts of CMB probes. We constrain the scale of non-locality and report the corresponding values of $(n_t,\,r)$.

Throughout this paper we use the units $M_p^2=1/8\pi G$ and use the metric signature $(-,+,+,+)$. Also everywhere over-dot denotes derivative with respect to cosmic time $t$.  

\section{From local to non-local Starobinsky inflation}

\label{gen-sec}

Let us briefly review here the Starobinsky theory or local $R^2$ gravity whose action is given by 

\begin{equation}
S_{R^2}=\int d^4x\sqrt{-g}\left[\frac{M_p^2}{2}R+\frac{M_p^2}{12M^2}R^2\right]\,.
\label{localR2}
\end{equation}
From the above action we can clearly read that the quadratic curvature term would become dominant when when the Ricci scalar takes the values $R\gg 6M^2$ with $M$ being the effective scale $M\sim 5.5\times 10^{-5} M_p$. The above action has been so far be the successful model of inflationary cosmology.  

The equations of motion of local $R^2$ gravity are 
\begin{equation}
E^\mu_\nu\equiv -\left(M_p^2+\frac{M_p^2}{3M^2}R\right)G^\mu_\nu-\frac{M_p^2}{12M^2}R^2\delta^\mu_\nu+\frac{M_p^2}{3M^2}\left(\nabla_\mu\nabla_\nu-\delta^\mu_\nu\Box\right)R=0\,.
\label{R2INF}
\end{equation} 

The trace equation is 

\begin{equation}
\Box R=M^2R\,.
\label{Staro}
\end{equation}
Considering Friedmann-Lama\^itre-Robertson-Walker (FLRW) metric 

\begin{equation}
ds^2=-dt^2+a(t)^2d\boldsymbol{x}^2\,,
\end{equation}
solving the $00$ equation with quasi-dS approximation i.e. $\epsilon=-\frac{\dot{H}}{H^2}\ll 1$ gives the following scale factor solution 

\begin{equation}
\begin{aligned}
a&\approx a_0 (t_s-t)^{-\frac{1}{6}}e^{-\frac{M^2(t_s-t)^2}{12}}\,,\\
H=\frac{\dot{a}}{a}&\approx \frac{M^2}{6}(t_s-t)+\frac{1}{6(t_s-t)}\,,\\
R=12H^2+6\dot{H}&\approx \frac{M^4(t_s-t)^2}{3}-\frac{M^2}{3}+\frac{4}{(t_s-t)^2}\,.
\end{aligned}
\label{scale-fac}
\end{equation}

The solution (\ref{scale-fac}) is an approximate solution of $00$ equation (\ref{R2INF}) in the quasi-dS  limit. The interesting thing in the solution (\ref{scale-fac}) is that it gives an accelerated expansion when $t\ll t_s$ and it it ends at $t\sim t_s\sim \frac{1}{M}$. This is called graceful exit from inflation. Moreover, it was shown that solution of $00$ equation in the limit $t\gg t_s$ leads to matter dominated era \cite{Starobinsky:1980te} given by 

\begin{equation}
a\sim a_1t^{2/3}\LT 1+\frac{2}{3Mt}\sin M(t-t_1) \RT\,.
\end{equation}

Therefore, $R^2$ inflation nicely interpolates between inflation and matter dominated eras. Note that the solution (\ref{scale-fac}) is non-singular in the limit $t\to -\infty$. 

We can also numerically solve the $00$ equation (\ref{scale-fac}) with the initial conditions of quasi-dS regime which we get as shown in the figure below 

\begin{figure}[h]
	\centering\includegraphics[height=2.5in]{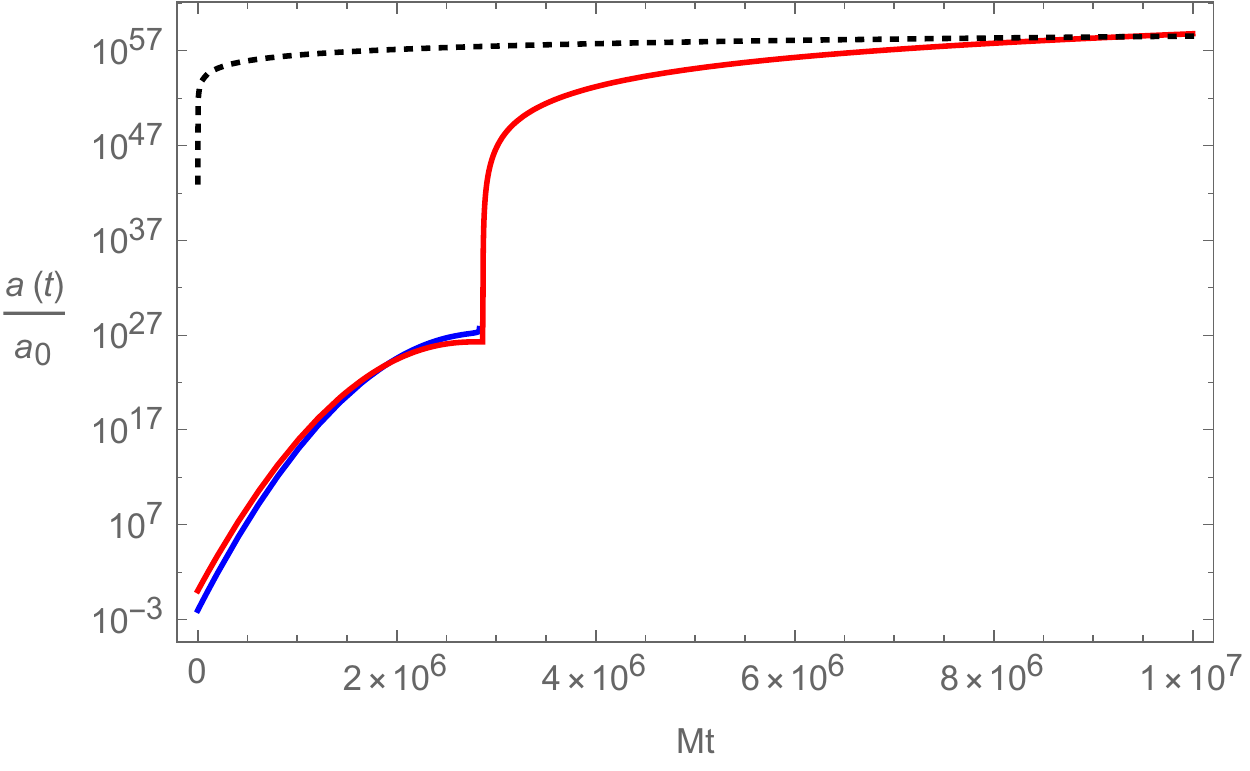}\caption{In the plot, the red line is the full numerical solution of $00$ equation of Starobinsky model. We chose initial conditions corresponds to the 60 $e$-folds and $M\sim 10^{-5}M_p$. The blue line is the approximate solution (\ref{scale-fac}) in the quasi-dS regime where we took $t_s=2.86\times 10^6 M_p^{-1}$. The dotted line is the scale factor $a\sim t^{2/3}$ which coincides with the full numerical solution at scales $t\gg \frac{1}{M}$.}	
	\label{Staro-plot} 
\end{figure}

Let us consider the non-local extension of Starobinsky theory \cite{Koshelev:2016xqb,Koshelev:2017tvv}

\begin{equation}
S=\int d^{4}x\sqrt{-g}\left[\frac{M_{p}^{2}}{2}R+\frac{1}{2}R\Fc_R\left(\frac{\Box}{\Mc^2}\right)R+\frac{1}{2}W_{\mu\nu\rho\sigma}\Fc_{W}\LF \frac{\Box}{\Mc^2}\RF W^{\mu\nu\rho\sigma}\right]\,.\label{NC-action}
\end{equation}
where the form factors are analytic functions of d'Alembertians 
\[\Fc_{R}\left(\frac{\Box}{\Mc^2}\right)=\sum_{n=0}^{\infty}f_{1n}\frac{\Box^n}{\Mc^{2n}}\,, \quad \Fc_{W}\left(\frac{\Box}{\Mc^2}\right)=\sum_{n=0}^{\infty}f_{2n}\frac{\Box^n}{\Mc^{2n}}\,.\]
Here $\Mc\lesssim M_p$ is called scale of non-locality.  The action (\ref{NC-action}) is found to be ghost free, superrenormalizable or finite and also is proved to be the most general action of gravity around MSS \cite{Koshelev:2016xqb,Koshelev:2017ebj,Koshelev:2017tvv}. 

Equations of motion (EoM) of this theory are given by

\begin{equation}
\begin{aligned}E_{\nu}^{\mu}\equiv & -\left[M_{p}^{2}+2\Fc_R\left(\frac{\Box}{\Mc^2}\right)R\right]G_{\nu}^{\mu}-\Lambda_{\rm cc} \delta^{\mu}_{\nu}-\frac{1}{2}R\Fc_R\left(\frac{\Box}{\Mc^2}\right)R\delta_{\nu}^{\mu}+2\left(\nabla^{\mu}\partial_{\nu}-\delta_{\nu}^{\mu}\Box\right)\Fc_R\left(\frac{\Box}{\Mc^2}\right)R\\
& +\mathcal{K}_{\nu}^{\mu}-\frac{1}{2}\delta_{\nu}^{\mu}\left(\mathcal{K}_{\sigma}^{\sigma}+\bar{\mathcal{K}}\right)+2(R_{\alpha\beta}+\nabla_\alpha\nabla_\beta)\Fc_W\left(\frac{\Box}{\Mc^2}\right)W_\nu^{\alpha\beta\mu}+\Oc\LF W^2 \RF=0\,.
\end{aligned}
\label{EoM}
\end{equation}
where $\Oc\LF W^2 \RF$ terms can be read from \cite{Biswas:2013cha} and

\[
\begin{aligned}\mathcal{K}_{\nu}^{\mu}= \frac{1}{\Mc^2} \sum_{n=1}^{\infty}f_{1n}\sum_{l=0}^{n-1}\partial^{\mu}\frac{\Box^{l}}{\Mc^{2l}}R\partial_{\nu}\left(\frac{\Box}{\Mc^2}\right)^{n-l-1}R\,,\quad
\bar{\mathcal{K}}=  \sum_{n=1}^{\infty}f_{1n}\sum_{l=0}^{n-1}\frac{\Box^{l}}{\Mc^{2l}}R\left(\frac{\Box}{\Mc^2}\right)^{n-l}R\,.
\end{aligned}
\]

The trace equation is  

\begin{equation} 
E=\left[M_p^2 -6\Box\Fc_{R}\left(\frac{\Box}{\Mc^2}\right) \right]R-\Kc_\mu^\mu-2\Kc+\Oc\LF W^2\RF=0\,.
\label{tEoMtrace}
\end{equation}

By assuming the trace equation of Starobinsky theory implies a recursive relation upon the action of infinite d'Alembertians in the form factor $\Fc_R(\Box/\Mc^2)$

\begin{equation}
\begin{aligned}
\Box R=M^2R \implies \Fc_R\LF\frac{\Box}{\Mc^2}\RF R=\Fc_1R\,,
\end{aligned}
\label{non-locansatz}
\end{equation}
from which it follows that the trace equation becomes 

\begin{equation}
\LF M_p^2-6M^2\Fc_1\RF R-\frac{1}{\Mc^2}\Fc_R^{(1)}\LF\frac{M^2}{\Mc^2}\RF\LF \pd_\mu R\pd^\mu R+2M^2R^2\RF=0\,.
\end{equation}
which gives the following unique conditions on the form factor\footnote{We are only interested here FLRW spacetime as it is conformally flat Weyl tensor contributions to the background equations vanish here.} (if $R(t)\neq 0$).

\begin{equation}
\Fc_1=\Fc_R\LF\frac{M^2}{\Mc^2}\RF=\frac{M_p^2}{6M^2}\,\quad \quad \Fc_R^{(1)}\LF\frac{M^2}{\Mc^2}\RF=0\,.
\label{condiFR}
\end{equation}
where $^{(1)}$ derivative with respect to the argument. We can notice that $\Fc_1$ is exactly same the the coefficient of $R^2$ term in the Starobinsky theory (\ref{localR2}). It was shown in \cite{Koshelev:2017tvv} that the trace equation of Starobinsky theory happens to be the only solution of (\ref{tEoMtrace}). Therefore, Starobinsky inflation is an attractor in non-local gravity. 

The conditions on the form factor (\ref{condiFR}) are quite imporant here as we will see in the next section the derivation of form factors. 

\section{UV completing Starobinsky theory}

\label{sec.starouv}

As we saw in the previous section, Starobinsky solution happens to be more closer and closer to exact dS in the limit $t\to -\infty$. This can be understood as the curvature $R\ggg 6M^2$ the quadratic term tend to dominate and we enter into almost scale invariant gravity. 
 Tt was recently shown the Starobinsky theory (local $R^2$) might suffer with a Benlinsky-Kalatnikov-Lifnitz (BKL) type of singularity as a past attractor to inflationary phase \cite{Muller:2017nxg}, however, in the non-local version in Weyl basis (\ref{NC-action}) we can trivially avoid the BKL type singularities \cite{Koshelev:2018rau}. So Starobinsky theory can be UV complete in non-local theory in Weyl basis if we can find form factors which must analytic and leads to no ghosts in near dS backgrounds. In \cite{Koshelev:2016xqb} the structure of form factors satisfying (\ref{condiFR}) found only for Minkowski. In \cite{Biswas:2016egy,Koshelev:2017ebj} form factors were found only for exact dS and Anti-dS which are not exactly analytic (or can be analytic with certain assumptions). Here we compute more carefully within near dS approaximation the form factors that obey (\ref{condiFR}) and prove that the theory must contain a scalaron responsible for inflationary perturbations.
 
 The full second order perturbed action with the form factor satisfying $\Fc_R^{(1)}\LF\frac{M^2}{\Mc^2}\RF=0$ around any solution of (\ref{non-locansatz}) can be found as \cite{Koshelev:2016xqb} 
 
 \begin{equation}
 \begin{aligned}
 \delta^2S=\int d^4x\sqrt{-g}\bigg[ &\delta_{local}+\zeta\Zc\LF\frac{\Box}{\Mc^2}\RF\zeta +\delta W_{\mu\nu\rho\sigma}\Fc_W\LF\frac{\Box}{\Mc^2}\RF \delta W^{\mu\nu\rho\sigma}\bigg]\,,
 \end{aligned}
 \label{fullpert}
 \end{equation}
 where $\delta_{local}$ is the second variation of the local action 
 
 \begin{equation}
 \begin{aligned}
 S_{local}&=\int d^4x \sqrt{-g}\LT \frac{M_p^2}{2}R+\frac{\Fc_1}{2}R^2\RT\,
 \end{aligned}
 \end{equation} 
 which reads as 
 \begin{equation}
 \delta_{local}= \LT\frac{M_p^2}{2}+\Fc_1\bar{R}\RT\delta_0-\Fc_1\bar{R}^2\delta_g+\Fc_1(\delta R)^2\,,
 \label{local-part}
 \end{equation}
 here $\delta_0$ is the variation of Einstein-Hilbert part, $\delta_g=\delta^2(\sqrt{-g})$ and 
 \begin{equation}
 \zeta=\delta\Box R+(\bar{\Box}-r_1)\delta R\,,\quad \Zc\LF\frac{\Box}{\Mc^2}\RF=\frac{\Fc\LF\frac{\Box}{\Mc^2}\RF-\Fc_1}{\LF\Box-r_1\RF^2}\,.
 \end{equation}

We can easily understand the second term in the second variation (\ref{fullpert}) comes from the non-local quadratic term. As we know from Starobinsky solution (\ref{scale-fac}) that the Universe becomes very mush dS like especially for $t\ll t_\ast$. Therefore, to make the Starobinsky theory UV complete in non-local gravity it is enough to find suitable form factors around dS i.e., $\bar{R}\approx\const$ and $\bar{R}_{\mu\nu}\approx\frac{\bar{R}}{4}g_{\mu\nu}$. Therefore, for near dS spacetime we can approximate $\zeta\approx(\bar{\Box}-M^2)\delta R$. With this the second order perturbed action would become 

\begin{equation}
\delta^2S=\int d^4x\sqrt{-g}\LT \Fc_1(\bar{R}+3M^2)\delta_0-\Fc_1\bar{R}^2\delta_g+\delta R\Fc_R\LF\frac{\bar{\Box}}{\Mc^2}\RF\delta R+\delta W_{\mu\nu\rho\sigma}\Fc_W\LF\frac{\Box}{\Mc^2}\RF W^{\mu\nu\rho\sigma}\RT\,,
\label{dSact}
\end{equation}
where we made a substitution $M_p^2=6\Fc_1M^2$ from (\ref{condiFR}). The resultant action around dS (\ref{dSact}) should not lead to any ghost degrees of freedom. In other words, the form factors has to be such that they should be analytic and must satisfy the conditions (\ref{condiFR}) and should avoid ghost propagating degrees of freedom. We can find such form factors by expanding the action (\ref{dSact}) in terms of 4-dimensional scalar and tensor perturbations which can be decomposed as \cite{Biswas:2016egy,Koshelev:2017ebj}

\begin{equation}
g_{\mu\nu}\to \bar{g}_{\mu\nu}+h_{\mu\nu}\,\quad h_{\mu\nu}\to h^\perp_{\mu\nu}-\frac{\phi}{4}\bar{g}_{\mu\nu}\,,
\end{equation}
where $\phi$ and $h_{\mu\nu}^\perp$ are the scalar and transverse and traceless tensor. 

Writing the scalar part of perturbed action (\ref{dSact}) is \cite{Biswas:2016egy,Koshelev:2017ebj}

\begin{equation}
\delta^2S_0=-\frac{1}{2}\int d^Dx\sqrt{-\bar{g}}\tilde{\phi}\Bigg\{\LF\bar{\Box}+\frac{\bar{R}}{3}\RF\LT 2\Fc_1\bar{R}-2(3\bar{\Box}+\bar{R})\Fc_R\LF\frac{\bar{\Box}}{\Mc^2}\RF\RT+6\Fc_1M^2\bar{\Box}\Bigg\}\tilde{\phi}\,,
\label{scalar-part}
\end{equation}
where $\tilde{\phi}=\sqrt{\frac{3}{32}}\phi$\,. We already learned that the quadratic curvature term become important only when $\bar{R}\gg 6M^2$. Therefore, the last term in  (\ref{scalar-part}) can actually be neglected. In this approximation we can actually rewrite the action (\ref{scalar-part}) in the following form 

\begin{equation}
\delta^2S_0=-\frac{1}{2}\int d^Dx\sqrt{-\bar{g}}\tilde{\phi}\Bigg\{\LF\bar{\Box}+\frac{\bar{R}}{3}\RF\LT 2\Fc_1\LF\bar{R}+3M^2\RF-2(3\bar{\Box}+\bar{R})\Fc_R\LF\frac{\bar{\Box}}{\Mc^2}\RF\RT\Bigg\}\tilde{\phi}\,,
\label{scalar-part-1}
\end{equation}

Thus, we have a pole in the propagator at $\bar{\Box}=\frac{\bar{R}}{3}$ which corresponds to the scalar of GR which is constrained. The remaining term depending on $\Fc_R\LF\frac{\Box}{\Mc^2}\RF$ should either give a no pole or a single pole. If there is no extra pole the theory become very much the same as GR and then we cannot realize exact Starobinsky model which requires a scalaron to explain CMB powerspectra \cite{Koshelev:2017ebj}. We show below that it is impossible to have taking into account the conditions on the form factor (\ref{condiFR}). 

Let us consider 
\begin{equation}
2\Fc_1\LF\bar{R}+3M^2\RF-2(3\bar{\Box}+\bar{R})\Fc_R\LF\frac{\bar{\Box}}{\Mc^2}\RF=e^{H_0\LF\frac{\bar{\Box}}{\Mc^2}\RF}\,,
\end{equation}
where $H_0$ is an entire function which must be valid on the whole complex plane. This gives no extrapoles in the propagator. Substituting $\Fc_1=\frac{M_p^2}{6M^2}$ from (\ref{condiFR}) and evaluating the above equation at $\bar{\Box}=M^2$ we obtain 

\begin{equation}
e^{H_0\LF\frac{M^2}{\Mc^2}\RF}=0\,.
\end{equation}
It is trivial to see the above condition is impossible to be valid by definition.  

Therefore, only way for us to make the theory ghost free is to have an extra pole in the scalar propagator. For this we require

\begin{equation}
2\Fc_1(\bar{R}+3M^2)-2(3\bar{\Box}+\bar{R})\Fc_R\LF\frac{\bar{\Box}}{\Mc^2}\RF=\Fc_1(\bar{R}+3M^2)\LF 1-\frac{\bar{\Box}}{M^2}\RF e^{H_0\LF\frac{\bar{\Box}}{\Mc^2}\RF}\,, 
\label{formfacold}
\end{equation}

Solving for the form factor from the above relation we can easily deduce that it cannot be analytic. Then we have assume the theory has to be valid only in some energy regime as in \cite{Biswas:2016egy} which implies the form factor is not fully analytic or we should take $\Fc_1=0$ as in  \cite{Koshelev:2017ebj} which does not satisfy (\ref{condiFR}). As a way out, we found here a very simple resolution to this problem and we can indeed make the form factor analytic. Instead of the form (\ref{formfacold}) we can have the following which also give exactly the pole at $\Box=M^2$ in the scalar propagator.

\begin{equation}
2\Fc_1(\bar{R}+3M^2)-2(3\bar{\Box}+\bar{R})\Fc_R\LF\frac{\bar{\Box}}{\Mc^2}\RF=\Fc_1(\bar{R}+3M^2)\LF 1-\frac{3\bar{\Box}+\bar{R}}{3M^2+\bar{R}}\RF e^{H_0\LF\frac{\bar{\Box}}{\Mc^2}\RF}\,.
\label{formfacnew}
\end{equation}
Now we can find the form factor as 

\begin{equation}
\Fc_R\LF\frac{\Box}{\Mc^2}\RF=\Fc_1(\bar{R}+3M^2)\LT  \frac{1-\LF1-\frac{3\bar{\Box}+\bar{R}}{3M^2+\bar{R}}\RF e^{H_0\LF\frac{3\bar{\Box}+\bar{R}}{\Mc^2}\RF}}{3\bar{\Box}+\bar{R}} \RT\,.
\label{FRnew}
\end{equation}
We can verify that the above form factor is analytic and satisfy (\ref{condiFR}) with a simple condition 

\begin{equation}
\Fc_R^{(1)}\LF \frac{M^2}{\Mc^2}\RF \implies e^{H_0\LF\frac{M^2}{\Mc^2}\RF}=1\,.
\label{H0r1}
\end{equation} 

Similarly, the tensor part of the action reads as 

\begin{equation}
\begin{aligned}
\delta^2S_2=&\frac{1}{4}\int d^4x\sqrt{-\bar{g}}h^{\perp}_{\mu\nu}\bigg\{ \LF\bar{\Box}-\frac{\bar{R}}{6}\RF\LT \Fc_1\bar{R}+\LF\Box-\frac{\bar{R}}{3}\RF\Fc_W\LF\bar{\Box}+\frac{\bar{R}}{3}\RF\RT+ 3\Fc_1M^2\bar{\bar{\Box}} \bigg\} h^{\perp{\mu\nu}}
\end{aligned}
\label{tenspart-1}
\end{equation}
We can rewrite (\ref{tenspart-1}) within the approximation $\bar{R}\gg 6M^2$ as  

\begin{equation}
\begin{aligned}
\delta^2S_2=&\frac{1}{4}\int d^4x\sqrt{-\bar{g}}h^{\perp}_{\mu\nu} \LF\bar{\Box}-\frac{\bar{R}}{6}\RF\LT \Fc_1(\bar{R}+3M^2)+\LF\bar{\Box}-\frac{\bar{R}}{3}\RF\Fc_W\LF\frac{\bar{\Box}}{\Mc^2}+\frac{\bar{R}}{3\Mc^2}\RF\RT h^{\perp{\mu\nu}}\,.
\end{aligned}
\label{tenspart-1}
\end{equation}
The pole at $\bar{\Box}=\frac{\bar{R}}{6}$ is the standard GR pole and we must have the form factor of the Weyl square term to be such that we do not have any extra poles. We can achieve this by the following demand 

\begin{equation}
\Fc_1(\bar{R}+3M^2)+\LF\bar{\Box}-\frac{\bar{R}}{3}\RF\Fc_W\LF\bar{\Box}+\frac{\bar{R}}{3}\RF=\Fc_1(\bar{R}+3M^2)e^{2H_2\LF\frac{\bar{\Box}-\bar{R}/3}{\Mc^2}\RF}\,,
\label{usehij}
\end{equation}
from which we can extract the form factor as 

\begin{equation}
\Fc_W\LF\frac{\Box}{\Mc^2}\RF=\Fc_1(\bar{R}+3M^2)\LT \frac{e^{2H_2\LF\frac{\bar{\Box}-2\bar{R}/3}{\Mc^2}\RF}-1}{\Box-\frac{2}{3}\bar{R}} \RT\,.
\label{FW-new}
\end{equation}
We can easily verify that the above form factor is also  analytic. 

The entire functions $H_{\ell}\,,\ell=0,2$ define the UV completion of the theory. These should satisfy the properties we listed in Appendix.~\ref{appA}.

\section{Inflationary predictions of non-local Starobinsky theory}

\label{sec.staroin}

Inflationary perturbations of non-local Starobinsky inflation has been studied in detail \cite{Craps:2014wga,Koshelev:2016xqb,Koshelev:2017tvv}. Here in this section we briefly review those computations in relation to our modified form factors (\ref{FRnew}) and (\ref{FW-new}). Inflationary perturbations are usually computed in 3+1 Arnott-Deser-Misner (ADM) decomposition of the metric which is given by 

\begin{equation}
ds^{2}=a^{2}\left(\tau\right)\left[-\left(1+2\Phi\right)d\eta^{2}+\left(\left(1-2\Psi\right)\delta_{ij}+2h_{ij}\right)dx^{i}dx^{j}\right]\,.\label{line-element}
\end{equation}
where $\tau$ is conformal time, $\Phi,\Psi$ are Bardeen potentials and $h_{ij}$ is transverse and traceless tensor perturbations. Working out the linearized equations of motion for the metric (\ref{line-element}) we obtain \cite{Koshelev:2016xqb} 

\begin{equation}
\LT \Fc_1(\bar{R}+3M^2)+\LF\bar{\Box}-\frac{\bar{R}}{6}\RF\Fc_W\LF\frac{\bar{\Box}+\frac{\bar{R}}{2}}{\Mc^2}\RF\RT\frac{\Phi+\Psi}{a^2}=0\,.
\end{equation}
Substituting (\ref{FW-new}) in the above relation we get 

\begin{equation}
\Fc_1(\bar{R}+3M^2)e^{H_2\LF\frac{\bar{\Box}-\frac{\bar{R}}{6}}{\Mc^2}\RF}\frac{\Phi+\Psi}{a^2}=0\,.
\end{equation}
The only solution of the above equation is $\Phi+\Psi=0$. Given this we can compute the scalar power spectrum of the canonical variable defined by $\Upsilon=\Fc_R\LF\frac{\Box}{\Mc^2}\RF\delta R=2\Fc_1\LF\bar{R}+3r_1\RF\Phi$ using the perturbed action in quasi-dS regime 

\begin{equation}
\delta^2S^{(s)}=\frac{1}{2}\int d^4x \sqrt{-\bar{g}} \LT \LF 2\delta_{EH}^{(s)}-\bar{R}\delta g^{(s)}\RF f_0\bar{R}+\Upsilon\frac{1}{\Fc_R\LF\frac{\Box_k}{\Mc^2}\RF}\Upsilon\RT\,,
\label{in-SA}
\end{equation}
where the superscript ${(s)}$ denotes the scalar part and 

\begin{equation}
\delta_{EH}^{(s)}=-\frac{6}{a^2}\LF \Phi^{\prime 2}+k^2\Phi^2\RF\,, \quad
\delta_g^{(s)}=4\Phi^2\,,\quad 
\delta R=-2\LF\bar{R}+3\bar{\Box}\RF\Phi\,.
\end{equation}
Here 
\begin{equation}
\Wc\LF\frac{\Box}{\Mc^2}\RF= 3\Fc_R\LF\frac{\Box}{\Mc^2}\RF+ \LF R+3M^2\RF\frac{\Fc_R\LF\frac{\Box}{\Mc^2}\RF-\Fc_1}{\Box-M^2}\,.
\end{equation}
Following (\ref{formfacnew}) we can easily deduce that 
\begin{equation}
\Wc\LF\frac{\Box}{\Mc^2}\RF=3\Fc_1 e^{H_0\LF\frac{\bar{\Box}}{\Mc^2}\RF}\,.
\end{equation}
Therefore this operator does not cause an extra pole in the propagator of canonical scalar perturbation $\Upsilon$. 

Working out (\ref{in-SA}) we obtain  

\begin{equation}
\delta^2S^{(s)}= \frac{1}{2\Fc_1\bar{R}}\int d\tau d^3x\sqrt{-\bar{g}}\Upsilon\frac{\Wc\LF\frac{\bar{\Box}}{\Mc^2}\RF}{\Fc_R\LF\frac{\bar{\Box}}{\Mc^2}\RF}(\bar{\Box}-M^2)\Upsilon\,.
\end{equation}
Redefining the canonically normalized scalar $u=\sqrt{\frac{\Wc\LF\frac{\Box}{\Mc^2}\RF}{2\Fc_1\bar{R}\Fc_R\LF\frac{\Box}{\Mc^2}\RF}}\Upsilon$ we obtain 

\begin{equation}
\delta^2 S^{(2)}= \frac{1}{2}\int d\tau d^3x\sqrt{-\bar{g}}u\LF \bar{\Box}-M^2\RF u\,.
\label{canuac}
\end{equation}
Defining $\tilde{u}=au$ we obtain 

\begin{equation}
\delta^2 S^{(2)}= \frac{1}{2}\int d\tau d^3x \LT (\tilde{u}^\prime)^2-(\nabla u)^2+\frac{z_s^{\prime\prime}}{z_s}\tilde{u}^2\RT \,, 
\end{equation}
where

\begin{equation}
\frac{z_s^{\prime\prime}}{z_s}=\frac{\nu_s^2-\frac{1}{4}}{\tau^2}= 2a^2H^2-a^2M^2\,, 
\end{equation}
with 
\begin{equation}
\begin{aligned}
\nu_s &= \sqrt{\frac{9}{4}-\frac{M^2}{H^2}}\,,\\
&\approx \sqrt{\frac{9}{4}-6\epsilon}.
\end{aligned}
\end{equation}

The equations of motion for the Fourier mode given by 

\begin{equation}
\tilde{u}(\tau,\,\boldsymbol{x})=\int \frac{d^3k}{(2\pi)^3} \tilde{u}_{\boldsymbol{k}}(\tau)e^{i\boldsymbol{k}.\boldsymbol{x}}
\end{equation}
is 
\begin{equation}
\tilde{u}_k^{\prime\prime}+\LF k^2-\frac{z_s^{\prime\prime}}{z_s}\RF \tilde{u}_k=0\,,
\end{equation}
whose general solution can be written in terms of Hankel functions \cite{DeFelice:2010aj} 

\begin{equation}
\tilde{u}_k= \frac{\sqrt{\pi}}{2} (-\tau)^{1/2} \LT c_{1k} H^{(1)}_{\nu_s}\LF -k\tau\RF+c_{2k} H^{(2)}_{\nu_s}\LF-k\tau\RF\RT\,,
\end{equation}

The Bunch-Davies Vacuum initial conditions are $c_{1k}=1,\,c_{2k}=0$ which correspond to quantizing the perturbations in the sub-horizon limit $\vert k\tau\vert \gg 1$. In the leading order in slow-roll, the sub-horizon limit of the mode function can be approximated as 

\begin{equation}
\tilde{u}_k = \frac{H}{\sqrt{2k^3}}e^{-ik\tau}(1+ik\tau)\,.
\label{tildeuk}
\end{equation}

The power spectrum is usually defined as the two point correlation function of the Bardeen potential $\Phi$ 

\begin{equation}
\langle 0\vert  \hat{\Phi}\LF \boldsymbol{x},\,\tau \RF \hat{\Phi}\LF \boldsymbol{x}+\boldsymbol{r},\,\tau \RF\vert 0\rangle = \int_0^\infty \frac{dk}{k} \frac{\sin kr}{kr}\Big\vert \delta_\Phi\LF \boldsymbol{k},\,\tau\RF \Big\vert^2\,. 
\end{equation}

We can compute $\Big\vert \delta_\Phi\LF \boldsymbol{k},\,\tau\RF \Big\vert^2$ as 

\begin{equation}
\begin{aligned}
 \Big\vert\delta_\Phi\LF \boldsymbol{k},\,\tau\RF \Big\vert^2=\frac{1}{4\pi^2} \frac{1}{ 2\Fc_1\bar{R}}\Big\vert \hat{\Upsilon}\LF \boldsymbol{k},\,\tau \RF \Big\vert^2 &= \frac{1}{4\pi^2} \frac{1}{3\Fc_1^2} \frac{1}{ 2\Fc_1\bar{R}} \Big\vert \Fc_R\LF\frac{\Box}{\Mc^2}\RF e^{-H_0\LF\frac{\Box}{\Mc^2}\RF} u_k\LF -k\tau\RF\Big\vert^2\,\\
 &= \frac{1}{16\pi^2} \frac{1}{ 2\Fc_1\bar{R}} \frac{1}{3} \Big\vert \frac{1}{a}\tilde{u}_k \Big\vert^2\,,
 \end{aligned}
 \label{powespec}
\end{equation}
where the factor of $4\pi^2$ comes from canonical quantization commutation relations \cite{Mukhanov:1990me}. In the above steps we used the fact that $u_k$ is eigen mode of the d'Alembertian with eigen value $M^2$ which can be read from (\ref{canuac}). We have also used (\ref{H0r1}). 

Substituting (\ref{tildeuk}) in (\ref{powespec}) and computing the power spectrum on super-horizon scales ($k\ll aH$) we obtain 

\begin{equation}
\vert \delta_\Phi(\boldsymbol{k})\vert^2=\frac{H^2}{16\pi^2}\frac{1}{3\Fc_1\bar{R}}\,.
\end{equation}


Now the primordial power spectrum ($\Pc_\Rc$) of the curvature perturbation $\Rc=-\Phi+H\frac{\delta R}{\dot{\bar{R}}}=-\Phi-\frac{24H^3}{24H\dot{H}}\Phi\approx -\frac{H^2}{\dot{H}} \Phi$ and its tilt can be computed as  \cite{Craps:2014wga,Koshelev:2016xqb} 

\begin{equation}
\left.\Pc_\Rc\right|_{k=aH}= \frac{H^2}{16\pi^2\epsilon^2}\frac{1}{3\Fc_1\bar{R}}\,,\quad  n_s\equiv \left.\frac{d\ln \Pc_\Rc}{d\ln k}\right|_{k=aH} \approx 1-\frac{2}{N} \,,\label{Psuni}\end{equation}
where $N$ is the number of $e$-folds and $\epsilon=-\frac{\dot{H}}{H^2}\approx\frac{1}{2N}$.

Similarly, using (\ref{usehij}) the second order action for the tensor perturbation is 

\begin{equation}
\begin{aligned}
\delta^2S_2=&\frac{1}{4}\int  d^4x\sqrt{-\bar{g}}h^{\perp}_{ij} \Fc_1\LF \bar{R}+3M^2\RF e^{2H_2\LF \frac{\Box-\frac{\bar{R}}{3}}{\Mc^2}\RF} \LF\bar{\Box}-\frac{\bar{R}}{6}\RF h^{\perp{ij}}\,.
\end{aligned}
\label{tenspart}
\end{equation}

Computing the tensor power spectrum is very similar to the scalar power spectrum. It was carried out in \cite{Koshelev:2016xqb,Koshelev:2017tvv}, here the results with respect to the form factor (\ref{FW-new}). The tensor power spectrum and the tilt can be straight forwardly computed as \cite{Koshelev:2017tvv}

\begin{equation}
\begin{aligned}
\left.\mathcal{P}_{\mathcal{T}}\right|_{k=aH}&=\frac{1}{12\pi^{2}\mathcal{F}_{1}}\left(1-3\epsilon\right)e^{-2H_2\left(-\frac{\bar{R}}{3\mathcal{M}^{2}}\right)}\,\\
n_{t}\equiv\left.\frac{d\ln\mathcal{P}_{\mathcal{T}}}{d\ln k}\right|_{k=aH}&\approx-\frac{d\ln\mathcal{P}_{\mathcal{T}}}{dN}\left(1+\frac1{2N}\right)\\
&\approx-\frac3{2N^2}-\left(\frac 1{2N}+\frac1{4N^2}\right)\frac{\bar{R}}{3\mathcal{M}^{2}}H_2^{\prime}\left(-\frac{\bar{R}}{3\mathcal{M}^{2}}\right)\,.
\end{aligned}
\label{Ttilt}
\end{equation}

The ratio of tensor power spectra to scalar power spectra is given by 

\begin{equation}
r=\frac{12}{N^2}e^{-2H_2\LF\frac{\bar{R}}{3\Mc^2}\RF}\,.
\label{T2S}
\end{equation}

\subsection{Form factor of Tomboulis (1997)}

Form factors define the UV properties of the theory such as (super)renormalizability. In the view of non-local Starobinsky inflation, given a suitable form factor we can indeed put some observational numbers to test the theory. In this respect, let us consider entire function proposed by Tomboulis \cite{Tomboulis:1997gg}

\begin{equation}
H_2(z)= \log \left(p(z)^2\right)+\gamma_E+\Gamma \left(0,p(z)^2\right)\,,
\label{Tomboulis}
\end{equation}
where $p(z)$ is the polynomial of degree $\gamma+1$ and $\gamma \geq 2$ is an integer.

\begin{equation}
p(z)= b_{\gamma+1}z^{\gamma+1}+b_{\gamma}z^{\gamma}+...+b_2z^2\,.
\label{pz}
\end{equation}
 For the inflationary predictions (\ref{Ttilt},\ref{T2S}) we evaluate the above entire function at $z_\ast=-\frac{\bar{R}}{6\Mc^2}$. Naively, we expect to change the predictions with respect to Starobinsky model only when\footnote{$z_\ast\to 0 \implies H_2(z_\ast)\to 0$ in which case deviation from the predictions of local $R^2$ model negligibly small to be probed by future observations.} $z_\ast=\Oc(1)$. Therefore, we take the range of non-locality to be  $M^2\ll \Mc^2 \lesssim \bar{R}$. For this range of non-local scale only the first two terms in the polynomial (\ref{pz}) are important. All the rest of the terms of order $\lesssim \gamma-1$ can be neglected as they do not effect greatly the predictions (\ref{Ttilt}) and (\ref{T2S}). Here below we plot tensor to scalar ratio (\ref{T2S}) and tensor tilt (\ref{Ttilt}) with respect to different values of $\gamma$.

\begin{figure}[H]
	\centering\includegraphics[height=2.4in]{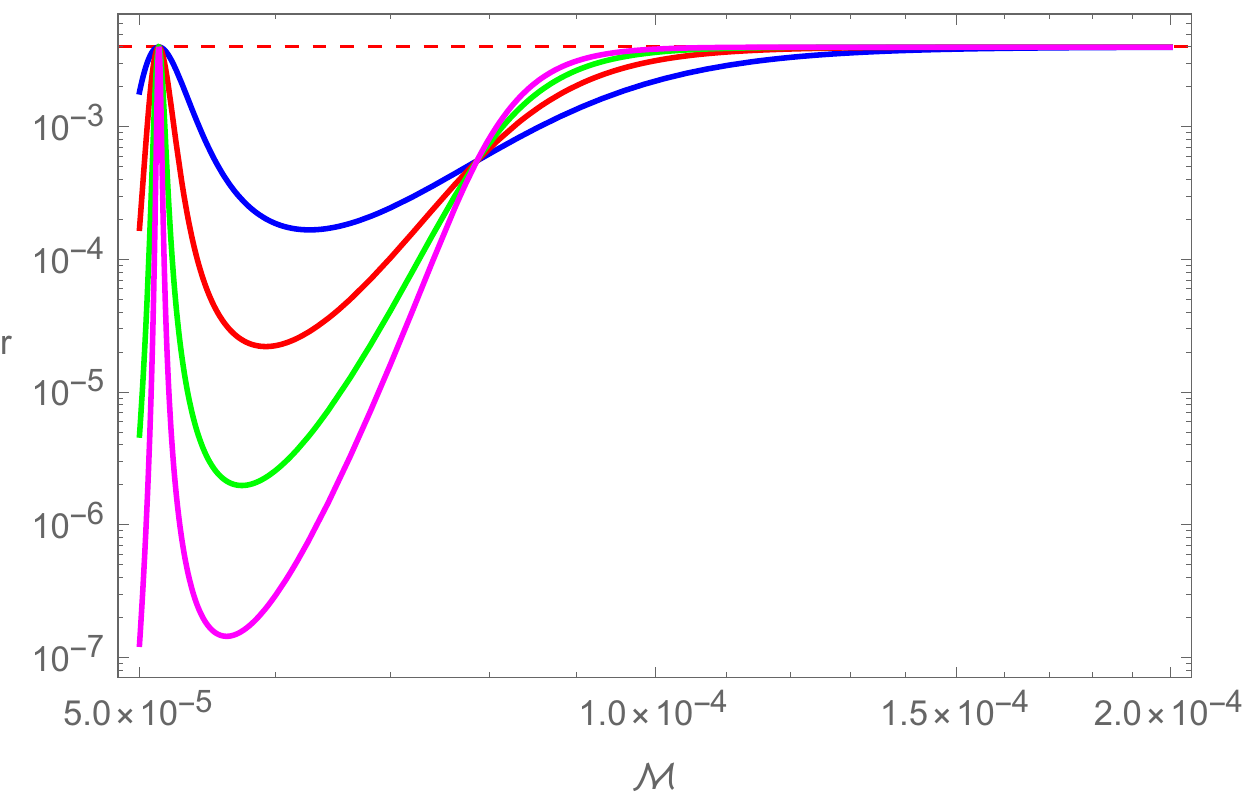}\quad \includegraphics[height=2.4in]{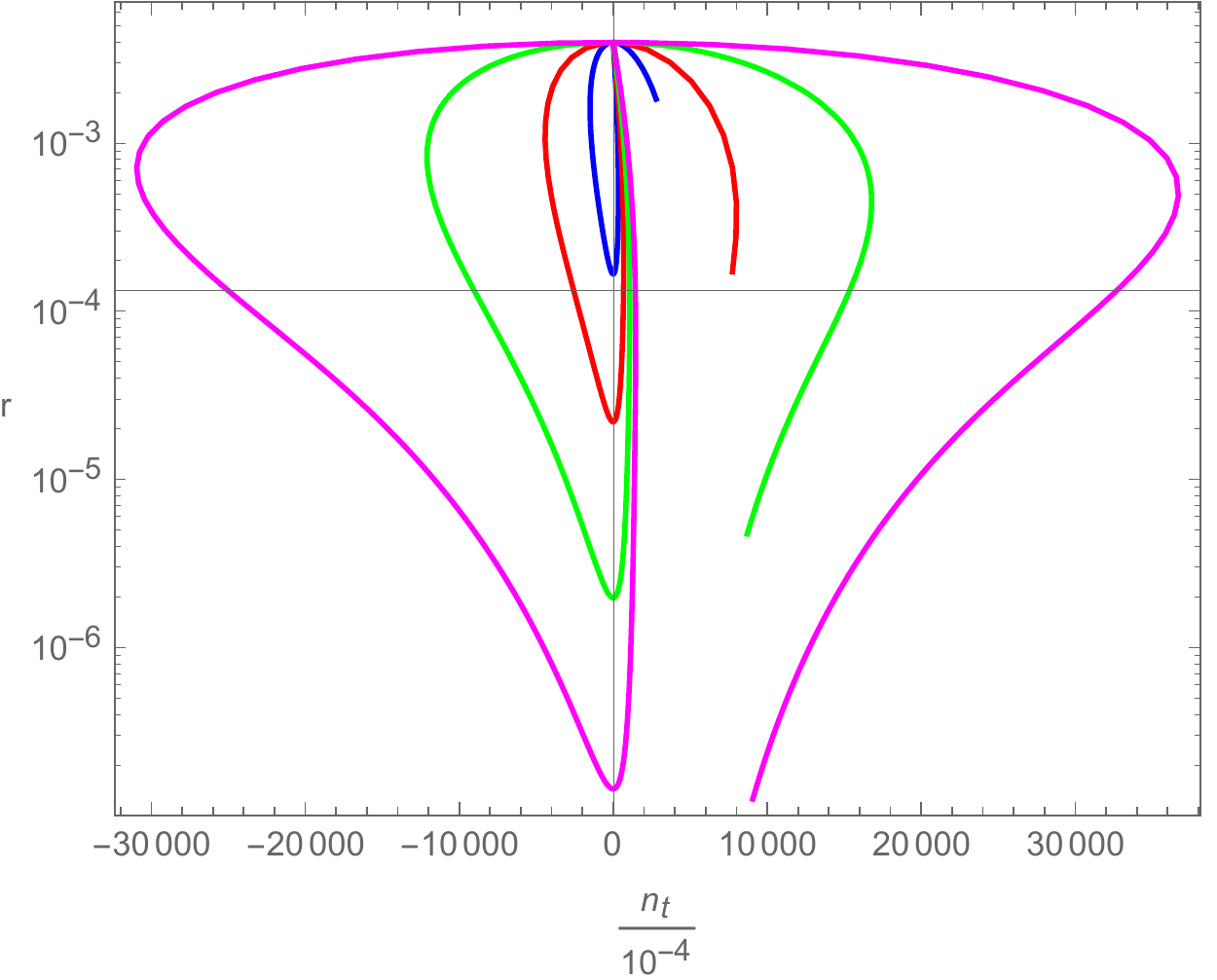}\caption{In the left panel we plot tensor to scalar ratio Vs. the scale of non-locality $\Mc$ (in the units of $M_p$). Right panel is the parametric plot of $n_t-r$ with $\Mc$. For both the plots we took $b_{\gamma+1}=0.85,\,b_{\gamma}=2$. The colored lines blue, red, green, magenta represent $\gamma=2,3,4,5$ respectively. The form factor we consider here is (\ref{Tomboulis}). We considered $N=55$.}
	\label{Staro-plot-1} 
\end{figure}
From the above plot we can easily notice that $r\lesssim 0.004$ and $n_t\lessgtr 0$ depending on the scale of non-locality and the value of $\gamma$. So the tensor to scalar ratio here is always below the value of Starobinsky model. 

To get tensor to scalar ratio above the values of Starobinsky model, we propose a more general entire function which still do not spoil the properties \ref{appA} in the limit $z\to \infty$.

\begin{equation}
H_2(z)= \log \LT \frac{p(z)^2}{g(z)^2}\RT+\gamma_E+\Gamma \left(0,p(z)^2\right)-\Gamma \left(0,g(z)^2\right)\,,
\label{Tomboulisnew}
\end{equation} 
where $g(z)$ is the polynomial of degree less than $\gamma+1$. 

\begin{figure}[H]
	\centering\includegraphics[height=2.3in]{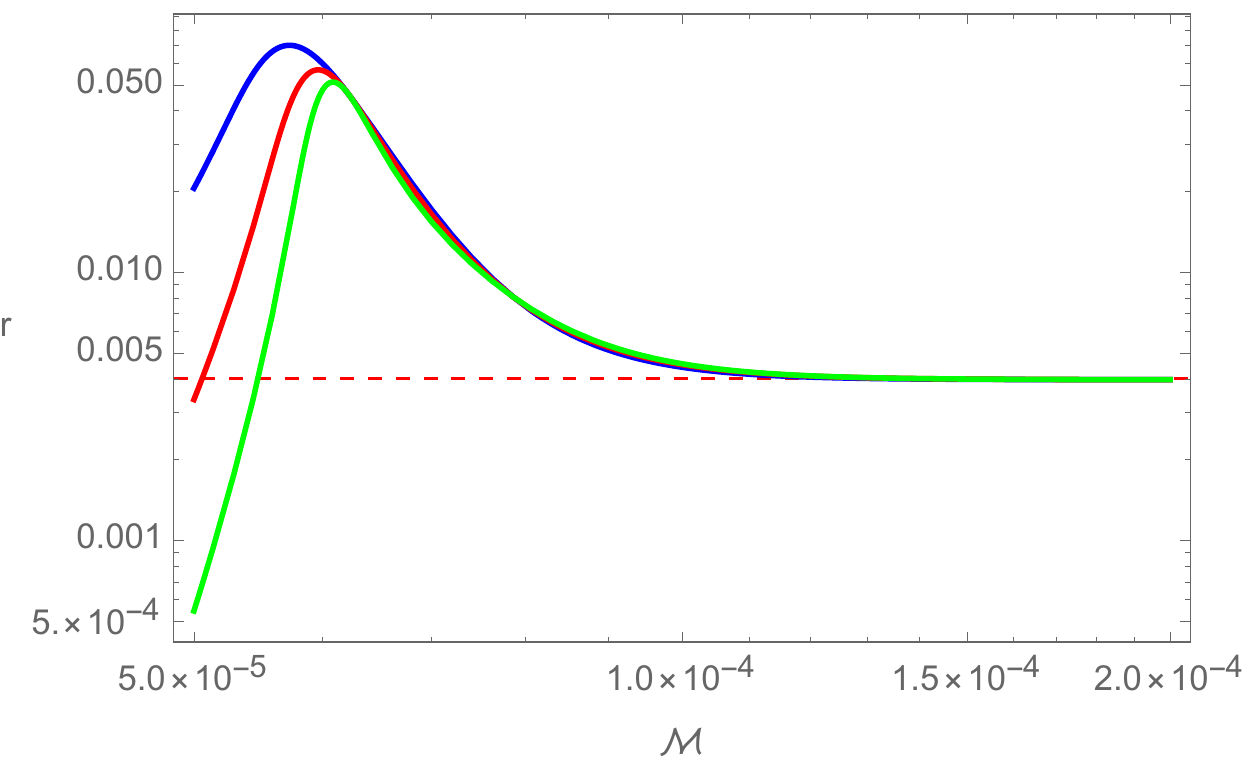}\quad \includegraphics[height=2.3in]{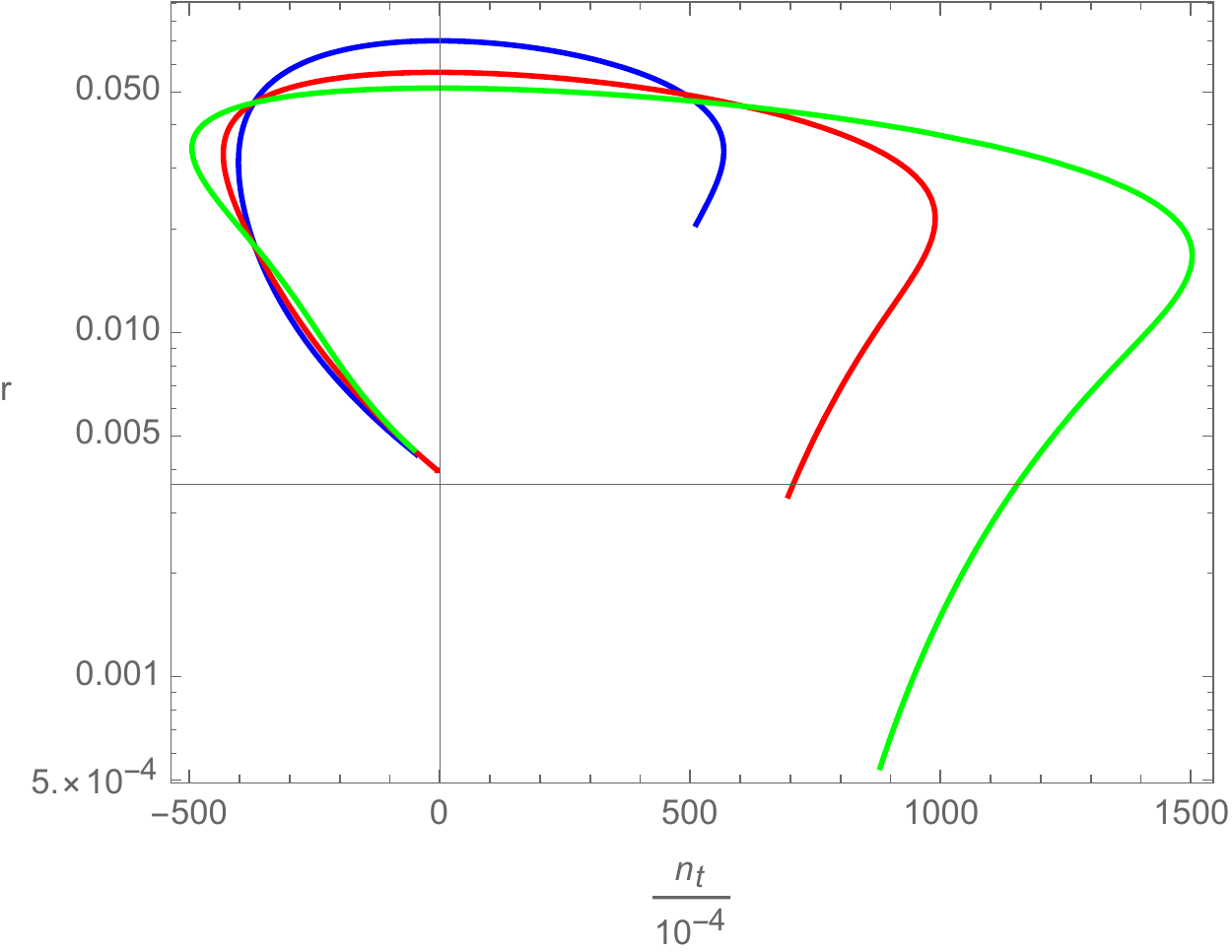}\caption{In the left panel we plot tensor to scalar ratio Vs. the scale of non-locality $\Mc$ (in the units of $M_p$). Right panel is the parametric plot of $n_t-r$ with $\Mc$. For both the plots we took $b_{\gamma+1}=0.5,\,b_{\gamma}=0.8$. The colored lines blue, red, green, magenta represent $\gamma=2,3,4,5$ respectively. The form factor we consider here is (\ref{Tomboulisnew}) and we took $g(z)=z^2$. We considered $N=55$.}.
	\label{Staro-plot-2} 
\end{figure}

\section{Commensts on Higgs inflation in non-local theory}

\label{sec.higgs.in}

In the local theories, one of the best models observationally as good as $R^2$ inflation is Higgs inflation \cite{Bezrukov:2007ep}. The difference between these models is quite small and major difference only happens with reheating \cite{Bezrukov:2011gp} which is difficult to probe with any near future CMB observations \cite{Martin:2015dha,Martin:2016oyk}. The astonishing observational coincidence of these two models has been  explained in \cite{Kehagias:2013mya} which we review below briefly. 

The Starobinsky theory (\ref{localR2}) can be simply written with an auxiliary scalar as 

\begin{equation}
S_{R^2}=\int d^4x\sqrt{-g}\LT \left(\frac{M_p^2}{2}+f_0\varphi\right)R-\frac{f_0}{2}\varphi^2\RT\,.
\label{StaroEm}
\end{equation}
If we do a conformal transformation $g_{\mu\nu}\to (M_p^2+f_0\varphi)^{-1}g_{\mu\nu}$ followed by a field redefinition $\tilde{\varphi}\to \tilde{\varphi}=\ln \sqrt{\frac{3}{2}}\LF 1+f_0\frac{\varphi}{M_p^2} \RF$ we go to Einstein frame where we will have a minimally coupled scalar field with flat potential known as Starobinsky potential \cite{Mukhanov:1990me,DeFelice:2010aj} 

\begin{equation}
S_{R^2}=\int d^4x\sqrt{-g}\LT \frac{M_p^2}{2}R-\frac{1}{2}\pd_\mu\tilde{\varphi}\pd^\mu\tilde{\varphi}- \frac{3}{4}M^2M_p^2\LF 1-e^{-\sqrt{\frac{2}{3}}\frac{\tilde{\varphi}}{M_p}} \RF \RT\,.
\label{StaroEF}
\end{equation}

Let us consider the Higgs model described by the action of the following form 

\begin{equation}
S_{H}=\int d^4x\sqrt{-g}\LT \frac{M_p^2}{2}R+\xi H^\dagger HR-\pd^\mu H^\dagger\pd_\mu H-\frac{\lambda}{4}\LF H^\dagger H-v^2\RF^2\RT\,,
\label{HigssIn}
\end{equation}
where $H$ is the SM Higgs doublet and in the Unitary guage $H=\frac{h}{\sqrt{2}}$, $\xi\gg 1$ is the large non-minimal coupling relevant where Higgs field is far away from its vacuum expectation value (VEV) ($\sim 240\,Gev$) $h^2\gg v^2$. The energy scale of inflation is usually around GUT scale $10^{16} GeV$ at which (\ref{HigssIn}) can be seen as 

\begin{equation}
S_{h}=\int d^4x\sqrt{-g}\LT \frac{M_p^2}{2}R+\frac{1}{2}\xi h^2R-\frac{1}{2}\pd^\mu h\pd_\mu h-\frac{\lambda}{4} h^4\RT\,. 
\label{HigssIn1}
\end{equation}
Conformal transformation of the above action into Einstein frame shows that the canonical scalar field potential exactly matches with Starobinsky potential in (\ref{StaroEF}) when $h\gg \frac{M_p}{\sqrt{\xi}}$. The inflatonary expansion is nearly the same as the one in Starobinsky model and $\xi \sim 10^4$ is what is required to match with CMB observations in Higgs inflation. The remarkable coincidence of Starobinsky and Higgs inflation is qualitatively discussed in \cite{Kehagias:2013mya}. In the action (\ref{HigssIn1}) if we neglect the Higgs kinetic term which is almost zero during inflation (slow-roll), it matches with (\ref{StaroEm}) with the mapping $M^2=\frac{\lambda}{\xi^2}$ and $\varphi=h^2$. This clearly explains why Starobinsky and Higgs inflation gives nearly same predictions. Because of this coincidence a natural question arises if inflation is driven by scalaron of $R^2$ gravity or Higgs. There are interesting studies recently combining these two models in a multifield set-up \cite{Calmet:2016fsr,He:2018gyf}. Besides Higgs inflation being as good as $R^2$ model, there are issues such as vacuum instability due to the requirement of large non-minimal coupling. This is because the Higgs self coupling runs to be negative as a result it suffers in instability. Recently, it was found that this issue can be well addressed with non-local abelian Higgs model inspired from string field theory \cite{Ghoshal:2017egr,Hashi:2018kag}.  Given we found UV completion of Starobinsky inflation in non-local gravity it is natural ask the question of UV completion of SM Higgs and its effect during inflation. In fact, super renormalizable or finite theory of gravity coupled to matter (UV complete standard model) established in \cite{Modesto:2015lna,Modesto:2015foa}. Here we aim to argue that using non-local Higgs theory we can exponentially suppress the Higgs dynamics at inflationary scales and as a result we can end up with non-local $R^2$ inflation. 

Extending the non-local Higgs model in \cite{Hashi:2018kag} with non-minimal coupling to gravity as 

\begin{equation}
S_{NH}=\int d^4x\sqrt{-g}\LT \frac{M_p^2}{2}R+\xi H^\dagger HR-\pd^\mu H^\dagger e^{-\frac{\Box^2}{\Mc^4}}\pd_\mu H+\mu^2H^\dagger e^{-\frac{\Box}{\Mc^4}} H-\frac{\lambda}{4}H^4\RT\,,
\label{HigssIn1}
\end{equation}
We assume $\xi\gg 1$. If the non-local scale is much higher than inflationary scale (as we saw in the previous section) both the kinetic term and the mass term in the above action are exponentially suppressed (assuming Higgs field slow-rolls). Naively comparing local (\ref{HigssIn}) and non-local Higgs theory (\ref{HigssIn1}) we can easily understand that the kinetic term of Higgs in non-local theory is heavily suppressed under standard slow-roll conditions. The remaining terms in (\ref{HigssIn1}) exactly reduces to local $R+R^2$ gravity. As we have seen in the previous section, $R^2$ model is UV complete with non-locality. Therefore, we expect in the context of non-local theories, $R^2$ is more natural to happen than Higgs inflation.

\section{Conclusions}

\label{conl}

The most general theory of gravity around MSS was shown to contain only the terms quadratic in Ricci scalar and Weyl tensor with AID non-local operators inserted in between. It has been proved that the space of solutions of this theory for conformally flat background are the same as those of local $R^2$ gravity. Therefore, the Starobinsky inflationary solution remains an attractor in this model. From the previous studies it is known that the theory in Weyl basis can be defined with an extra propagating massive scalar or without in addition to the usual massless graviton \cite{Biswas:2016egy,Koshelev:2017ebj}.  In this paper, using the most general solution of the theory for conformally flat backgrounds \cite{Koshelev:2017tvv} we find the general structure of form factors in near dS approximation and prove that the perturbative spectrum of AID non-local gravity in Weyl basis must contain a scalar degree of freedom (scalaron) , whose quantum fluctuations during inflation seed large scale structure formation. 

The presence of non-local Weyl square term leads to the modification of tensor spectrum as a result tensor to scalar ratio and the tensor spectral index gets modified compared to local $R^2$ model while the scalar spectral index remains the same \cite{Koshelev:2017tvv}. Considering the form factors which are known to give (super) renormalizability in the UV regime of the theory proposed by Tombounlis \cite{Tomboulis:1997gg} we study the observables $\LF r,\,n_t\RF$ and discuss the scale of non-locality $\Mc$ in the scope of observational signatures from the future CMB probes. With the form factor proposed in \cite{Tomboulis:1997gg} we obtain it is possible achieve tensor to scalar ratio as low as $10^{-7}\lesssim r\lesssim 3.3\times 10^{-3}$ depending on the scale of non-locality. Also we plot $\LF n_t,\,r\RF$ in Fig.~\ref{Staro-plot-1} where we can see possible to have both blue tilt and red tilt for tensor power spectrum being sensitive to the scale of non-locality. We also propose a generalization of the form factor to \cite{Tomboulis:1997gg} in (\ref{Tomboulisnew}) using which we obtain the predictions $5\times 10^{-4}< r\lesssim 0.07$ as a function of non-locality scale $ 5\times 10^{-5} <\frac{\Mc}{M_p}< 10^{-4}$. In Fig.~\ref{Staro-plot-2} we plot $\LF n_t,\,r\RF$ plane for this new form factor (\ref{Tomboulisnew}). Our results also show that in non-local Starobinsky model we can obtain either blue tensor tilt or red tensor tilt which can possibly be tested with future CMB probes such as CMBPol, COrE, LiteBIRD etc and several ground based detectors \cite{Creminelli:2015oda,Errard:2015cxa,Finelli:2016cyd}.

\appendix

\section{Properties of form factors}

\label{appA}

The entire functions $V^{-1}_{\ell}(z) \equiv \exp (H_{\ell}(z))$ ($z \equiv - \Box_{\Mc} \equiv - \Box/\Mc^2$) (for $\ell=0,2$) should 
satisfy the following general conditions \cite{Tomboulis:1997gg}:
\begin{enumerate}
	\renewcommand{\theenumi}{(\roman{enumi})}
	\item 
	$V^{-1}_{\ell}(z)$ is real and positive on the real axis and it has no zeros on the 
	whole complex plane $|z| < + \infty$. This requirement implies that there are no 
	gauge-invariant poles other than the transverse massless physical graviton pole;
	\item
	$|V^{-1}_{\ell}(z)|$ has the same asymptotic behaviour along the real axis at $\pm \infty$; 
	\item 
	There exist $\Theta>0$, $\Theta<\pi/2$ and positive integer $\gamma$, such that asymptotically
	\be
	&& 
	|V^{-1}_{\ell}(z)| \rightarrow | z |^{\gamma +1},\,\, {\rm when }\,\, |z|\rightarrow + \infty 
	\quad {\rm with} 
	\quad 
	\gamma\geqslant 2 \,,
		\label{tombocond}
	\ee 
	for the complex values of $z$ in the conical regions $C$ defined by: 
	$$C = \{ z \, | \,\, - \Theta < {\rm arg} z < + \Theta \, ,  
	\,\,  \pi - \Theta < {\rm arg} z < \pi + \Theta\}.$$
\end{enumerate}
The last condition is necessary to achieve the maximum convergence of the theory in
the UV regime.  
The necessary asymptotic behaviour is imposed not only on the real axis, but also on the conical regions, that surround it.  
In an Euclidean spacetime, the condition (ii) is not strictly necessary if (iii) applies. 

\section{Equations of motion for AID non-local Higgs non-minimally coupled to AID non-local gravity}

In this section, we write down the equations of motion for non-local Higgs model non-minimally coupled to AID non-local gravity whose action is given by

\begin{equation}
S_{NH}= \int d^4x\sqrt{-g}\LT \frac{M_p^2}{2}R+\xi H^\dagger HR- H^\dagger  \tilde{\Cc}\LF \frac{\Box}{\Mc^2}\RF  H-\frac{\lambda}{4}H^4\RT\,,
\end{equation}
where $\tilde{\Cc}\LF \frac{\Box}{\Mc^2}\RF=-\Cc\LF \frac{\Box}{\Mc^2}\RF -\mu^2=\sum_n c_n\frac{\Box^n}{\Mc^{2n}}$\,.

\begin{equation}
\begin{aligned}
\xi H^\dagger H G^\mu_\nu+E^\mu_\nu+\xi \LF \nabla^\mu\pd_\nu H^\dagger H -\delta^\mu_\nu\Box H^\dagger H\RF &= \frac{1}{2} \sum_{n=1}^{\infty} c_n \sum_{l=0}^{n-1}\Bigg[ \pd^\mu \Box^lH^\dagger\pd_\nu\Box^{n-l-1}H+\pd_\nu\Box^lH^\dagger\pd^\mu\Box^{n-l-1} H\\ &-\delta^\mu_\nu\LF \pd^\rho\Box^lH^\dagger \pd_\rho\Box^{n-l-1} H+\Box^lH^\dagger \Box^{n-l}H\RF \\&-\frac{1}{2}\delta^\mu_\nu \LF H^\dagger \Cc\LF\frac{\Box}{\Mc^2}\RF H-V  \RF \Bigg]
\end{aligned}
\end{equation}

\bibliographystyle{utphys}
\bibliography{Staronli}

\end{document}